\documentclass[10pt,preprintnumbers,twocolumn, amsmath,amssymbre,fleqn]{revtex4}

\usepackage{amsmath}
\usepackage{amsfonts}
\usepackage{amssymb}
\usepackage[colorlinks,linkcolor=blue,citecolor=red]{hyperref}
\usepackage{graphicx}
\usepackage{hyperref}
\usepackage{subfigure}
\usepackage{graphicx}
\usepackage{dcolumn}
\usepackage{bm}
\usepackage{amsthm,amsfonts}
\usepackage{graphicx}
\usepackage{physics}
\usepackage{hyperref}
\usepackage{lscape} 
\usepackage{amsmath}
\usepackage{amssymb}
\usepackage{graphics}
\usepackage{psfrag}
\usepackage{mathtools}
\usepackage{appendix}
\usepackage{comment}
\graphicspath{{figs/}}

\newcommand{\nc}{\newcommand}
\nc{\ba}{\begin{eqnarray}}
\nc{\ea}{\end{eqnarray}}
\nc{\bfk}{\bf{k} }
\nc{\bfq}{\bf{q} }
\nc{\bfp}{\bf{p} }

\nc{\rc}{\textcolor[rgb]{1.00,0.00,0.00}}
\nc{\bc}{\textcolor[rgb]{0.00,0.07,1.00}}

\newcommand{\calR}{{\cal{R}}}

\newcommand{\calP}{{\cal{P}}}

\begin{document}

\title{ Non-Perturbative Hamiltonian and  Higher Loop Corrections in USR Inflation}

\author{Hassan Firouzjahi}
 \email{firouz@ipm.ir}

\author{Bahar Nikbakht}
 \email{bahar.nikbakht@ipm.ir}

 \affiliation{ School of Astronomy, Institute for Research in Fundamental Sciences (IPM),
  P. O. Box 19395-5746, Tehran, Iran}
\begin{abstract}

Calculating the action and the interaction Hamiltonian at higher orders in cosmological perturbation theory is a cumbersome task.  {We employ the formalism of EFT of inflation in the decoupling limit for single-field 
ultra slow-roll (USR) inflation and obtain a non-perturbative Hamiltonian in terms of the Goldstone field $\pi$}. To complete the dictionary, a non-linear relation between the curvature perturbations and $\pi$ is presented. {Using these results, we compute higher-order loop corrections in USR models with a sharp transition to the attractor phase, relevant for PBHs formation}. 
It is shown that in the idealized picture in which the transition from the USR phase to SR phase is instantaneous and sharp,  the loop corrections on long CMB scales increase rapidly with the number of loops $L$ and the setup  may go out of perturbative control.

\end{abstract}

\maketitle


{\bf{Introduction: }}
The simple models of cosmic inflation are based on the dynamics of a single field, the inflaton field, with a nearly flat potential that is minimally coupled to gravity \cite{Weinberg:2008zzc, Baumann:2009ds}. During inflation, the background undergoes a short period of accelerated expansion during which all observed large scale structures are generated from quantum fluctuations. Models of inflations generally predict that the primordial perturbations in cosmic microwave background (CMB) and large-scale structure are nearly scale invariant and Gaussian as seen in cosmological observations \cite{Planck:2018jri}.

To study the predictions of inflation, one employs the formalism of cosmological perturbation theory in which the matter fields, as well as the gravitational perturbations, are studied on a homogenous and isotropic FLRW background \cite{Kodama:1984ziu, Mukhanov:1990me}. 
To calculate the correlators associated with the cosmological observables such as the curvature perturbations one needs the interaction Hamiltonian to the desired orders. For example, to calculate the two-point function or the power spectrum we only need the quadratic action and the free Hamiltonian. To calculate the three-point function or bispectrum the cubic Hamiltonian is needed.  Due to the non-linear structure of gravity, calculating the cubic and higher-order interactions proved to be very challenging. The first complete analysis of cubic action for the single field slow-roll (SR) 
models were performed by Maldacena \cite{Maldacena:2002vr}. Calculating 
the quartic and  higher-orders  actions in perturbation theory is a cumbersome task,  see \cite{Jarnhus:2007ia, Arroja:2008ga} for earlier works on quartic action.

Recently, models of ultra-slow-roll (USR) inflation have attracted significant interests. The USR setup is mostly used to generate primordial black holes (PBHs) as the candidate for dark matter \cite{Ivanov:1994pa, Garcia-Bellido:2017mdw, Germani:2017bcs, Biagetti:2018pjj}, see also 
\cite{Khlopov:2008qy, Ozsoy:2023ryl, Byrnes:2021jka, Escriva:2022duf, Pi:2024jwt, Pi:2022ysn} for further reviews.  The unique feature of the USR setup is that the curvature perturbation grows on superhorizon scales so large over-density can be generated at some cosmological scales as the seeds of PBHs. However, Kristiano and Yokoyama \cite{Kristiano:2022maq} raised the concern that this setup may not be under perturbative control. More specifically, they argued that the one-loop corrections from the small USR modes can back-react on the long CMB scale perturbations, affecting the  curvature perturbation power spectrum significantly. This issue has attracted significant interest in recent literature which is still under debate \cite{Kristiano:2022maq, Kristiano:2023scm, Riotto:2023hoz, Riotto:2023gpm, Choudhury:2023vuj,  Choudhury:2023jlt,  Choudhury:2023rks, Choudhury:2023hvf, 
Choudhury:2024one, Choudhury:2024aji, Firouzjahi:2023aum, Motohashi:2023syh, Firouzjahi:2023ahg, Tasinato:2023ukp, Franciolini:2023agm, Firouzjahi:2023btw, Maity:2023qzw, Cheng:2023ikq, Fumagalli:2023loc, Nassiri-Rad:2023asg, Meng:2022ixx, Cheng:2021lif, Fumagalli:2023hpa,  Tada:2023rgp,  Firouzjahi:2023bkt, Iacconi:2023slv, Davies:2023hhn, Iacconi:2023ggt, Kristiano:2024vst, Kristiano:2024ngc, Kawaguchi:2024lsw, Braglia:2024zsl, Firouzjahi:2024psd, Caravano:2024moy, Caravano:2024tlp, Saburov:2024und, Ballesteros:2024zdp, Firouzjahi:2024sce,Sheikhahmadi:2024peu, Frolovsky:2025qre, Inomata:2024lud, Kawaguchi:2024rsv, Fumagalli:2024jzz}. 

A major difficulty in calculating the loop corrections in cosmological perturbations is to calculate the interaction Hamiltonians beyond the cubic order. The effective field theory (EFT) formalism of inflation \cite{Cheung:2007st, Cheung:2007sv} provides a successful 
route to handle this difficulty. Indeed, using the EFT formalism the cubic and quartic interaction Hamiltonians were constructed in \cite{Firouzjahi:2023aum}    to calculate the full one-loop corrections. The analysis of \cite{Firouzjahi:2023aum} confirmed the conclusion of 
\cite{Kristiano:2022maq}, indicating that the one-loop correction can get out of perturbative control. This conclusion is more pronounced in a setup in which the USR phase is attached to the final SR phase instantaneously  and the mode function approaches its attractor value quickly after the USR phase. 
In this paper, building upon the analysis of \cite{Firouzjahi:2023aum}, we calculate the Hamiltonian to an arbitrary order in perturbation theory in the USR model. As an application of  this general Hamiltonian,  
we calculate the  loop corrections at an arbitrary order in the setup considered in \cite{Kristiano:2022maq}. For further details on the technical analysis see the companion paper \cite{longpaper}.  \\

{\bf Setup:} Here we briefly review the setup of USR inflation, see also 
\cite{Firouzjahi:2023aum} for further details about the model. The USR phase is a period of inflation in which the inflaton potential is exactly flat, $V(\phi)=V_0$ and the inflaton velocity falls off  as $a(t)^{-3}$ in which $a(t)$ is the FLRW scale factor. As a result, the first SR parameter $\epsilon \equiv 
\frac{-\dot H}{H^2}$ falls of as $a(t)^{-6}$ in which $H$ is the Hubble rate during inflation which is very nearly constant.  
In addition, the second SR parameter defined via $\eta\equiv \frac{\dot \epsilon}{H \epsilon}$ has the large value $\eta=-6$ \cite{Kinney:2005vj} which is unlike the usual SR models. Since the parameter $\epsilon$ falls off exponentially during the USR phase, the comoving curvature perturbation $\calR \propto {1}/{\sqrt{\epsilon}}$ grows on superhorizon scales as $a(t)^3$. 

A central issue in models involving USR is how to terminate it and glue it to an attractor phase. The simplest picture is employed in \cite{Namjoo:2012aa} in which the transition from the USR phase to the final attractor phase happens instantaneously, with the assumption that the mode function freezes immediately after the USR phase. This picture was relaxed in 
\cite{Cai:2018dkf} in which the mode function may evolve for some time after the USR phase before reaching its attractor value. This is controlled by the sharpness (or the relaxation) parameter $h$ which is defined via 
$\eta = -6 - h \theta(\tau -\tau_e)$ in which $d\tau=dt/a(t)$ is the conformal time. 
Across the transition point $\tau=\tau_e$, this yields,      
\ba
\label{eta-jump}
\frac{d \eta}{d \tau} = - h \delta (\tau -\tau_e)  \, ,  \quad \quad  \tau_e^- < \tau < \tau_e^+ \, .
\ea 

For example, in the setup studied in \cite{Namjoo:2012aa}, $h \rightarrow -\infty$ while in the model considered in \cite{Kristiano:2022maq}, $h=-6$. 
We are interested in the setup with sharp transition in which $|h| >1$ so we can safely neglect the subleading SR corrections in our analytical analysis  \cite{Cai:2018dkf, Firouzjahi:2023aum}. \\

{\bf Non-Perturbative  Hamiltonian:}
Here we present our result of  interaction Hamiltonian  to all order 
in perturbation theory in the USR setup \cite{longpaper}. For this purpose, we employ the formalism of EFT of inflation  \cite{Cheung:2007st, Cheung:2007sv}. In the homogenous and time-dependent FLRW background, the time dependence of the inflaton field breaks the usual four-dimensional diffeomorphism invariance to a three-dimensional diffeomorphism invariance. In the EFT picture, one considers all interactions that are consistent with this three-dimensional diffeomorphism invariance in the comoving gauge where the inflaton perturbations are turned off initially. Upon writing the action containing all terms allowed by the symmetry, one can restore the total four-dimensional diffeomorphism invariance by introducing the  Goldstone field $\pi(x^\mu)$. In a sense, the Goldstone field captures the inflaton perturbation $\delta \phi$.  
The great advantage of EFT formalism appears when one works in  the 
decoupling limit where one can neglect the perturbations associated with the lapse and shift functions in the metric sector. The errors in neglecting these perturbations in our analysis are at the order of $\epsilon^2$  which is a very good approximation in USR setup where $\epsilon$ is small. We have verified the validity of decoupling limit in our analysis to seven orders in perturbations. But the structure of the constrained equations suggests that the decoupling limit is valid to all orders in perturbation theory  \cite{Behbahani:2011it, Green:2024hbw}. 
Note that the above discussions about the decoupling limit concerns 
discarding $O(\epsilon^2)$ terms compared to $O(\epsilon)$ terms in our USR setup. This conclusion is unaffected where $\eta \sim O(1)$ since $\eta$ does not modify the dynamics of  the lapse and shift functions.  

Starting with the action of  matter  in single field models with 
the standard kinetic energy ($c_s=1$)  in the decoupling limit and discarding the subleading ${\cal O}(\epsilon^2)$ corrections, the action takes the following simple form \cite{Firouzjahi:2025ihn, Creminelli:2024cge}, \footnote{After our work appeared in arXiv, 
we became aware of [71] who obtained the action (2) too. 
We thank S. Renaux-Petel for bringing Ref. [71] to our attention. } 
\ba
\label{action}
S = M_P^2 H^2 \int  d^4x  a(t)^3 \epsilon( t+ \pi)  \left( \dot{\pi}^2 - \partial_i \pi \partial^i \pi \right) + {\cal O}(\epsilon^2) ,
\ea
in which $M_P$ is the reduced Planck mass. In obtaining the above action we have discarded a total derivative term which has the form $\frac{d}{dt}{\big( g(t) f(\pi)\big)}$ in which $g(t)$ is a function of background \cite{longpaper}. It is important that the discarded total derivative term is not a function of  $\dot \pi$. It is shown in  \cite{Braglia:2024zsl} that a total derivative term which is only a function of  $\pi$ (and not $\dot \pi$) can be absorbed by a canonical transformation in the  phase space so they are harmless. 
 
The next step is to construct the interaction Hamiltonian from the action  
(\ref{action}). The conjugate momentum density associated with the field 
$\pi$ is  $\Pi  = -2M_P^2 a^3 \dot \pi $. 
Constructing the full Hamiltonian density $\mathcal{H} = \Pi \dot{\pi} - \mathcal{L}$ form the above conjugate momentum and then subtracting the Hamiltonian of the free theory, the interaction Hamiltonian density is obtained to be,  
\ba
\label{HI-eq2}
\mathcal{H}_I = M_P^2 H^2 a(t)^3 \hspace{-0.5cm} && \Big[\epsilon(t)^2  \Big(\frac{1}{\epsilon(t+ \pi )}   -\frac{1}{\epsilon(t)}\Big) \dot \pi^2  \nonumber\\  
   &+&\Big(  \epsilon(t+\pi) - \epsilon(t) \Big)  \partial_i \pi \partial^i \pi \Big]  + {\cal O}(\epsilon^2) ,
\ea
in which it is understood that $\pi$ and $\dot \pi$ are in the interaction picture, i.e. constructed from the mode functions of the free theory. 
The above Hamiltonian is valid for any single field model with $c_s=1$ in the decoupling  limit. It can be used for an extended USR or constant-roll setup or a  USR phase which is glued to the SR phase. In the latter situation, one has to be careful with the jump in $\eta $ as parametrized in Eq. (\ref{eta-jump}). 
The fact that a compact non-perturbative expression such as   Eq. (\ref{HI-eq2})   can be obtained for the Hamiltonian is a realization of the power of EFT formalism in the decoupling limit. 

As a non-trivial example of the applicability of Eq. (\ref{HI-eq2}), consider the case of the bulk of USR (or a general constant-roll) in which $\eta$ is constant with no discontinuity. This corresponds to the domain which does not include the boundary at $\tau=\tau_e$   where the USR stage is glued to the SR phase. In this domain, 
$\epsilon\propto e^{\eta H t}$ so $\epsilon(t+\pi)= \epsilon(t) e^{\eta H \pi}$. Using this relation in Eq. (\ref{HI-eq2}) and discarding ${\cal O}(\epsilon^2)$ terms we obtain,
\ba
\label{HI-eq3}
\mathcal{H}_I = M_P^2 H^2 a^3 \epsilon \Big[  \big(e^{-\eta H \pi}-1\big) \dot \pi^2 +  \big( e^{\eta H \pi}-  1 \big)  \partial_i \pi \partial^i \pi \Big] .
\ea

Eq.~\eqref{HI-eq3} is a non-perturbative expression for the interaction Hamiltonian in terms  of $\pi( x^\mu)$, valid to all orders in perturbation theory. It is surprising and quite interesting that an exact non-perturbative Hamiltonian can be obtained. This is a unique feature of the USR setup where the system is simple enough and the subleading corrections can be neglected safely.    It can be checked that Eq.~\eqref{HI-eq3}  reproduces the cubic and quartic Hamiltonians of \cite{Firouzjahi:2023aum} in the bulk.

The Hamiltonian \eqref{HI-eq2} is written in terms of the Goldstone field $\pi$.
However, the observable quantity is the curvature perturbation $\calR$ which is measured at the end of inflation. Therefore, we need a dictionary to translate 
$\pi$ field in terms of $\calR$ to an arbitrary order. Following \cite{Maldacena:2002vr}, this can be achieved by going from the comoving gauge in which $\pi=0$ and the spatial part of the metric taking the form $h_{ij}= a(t)^2 e^{2 \calR}\delta_{ij}$ into the spatially flat gauge with a new time coordinate 
$\tilde t$,  in which $\pi(\tilde t)\neq 0$ and  $h_{ij}= a(\tilde t)^2 \delta_{ij}$. Relating these two coordinates by a general time diffeomorphism transformation, one 
obtains the following relation \cite{longpaper},
\ba
\label{R-eq}
\calR = \sum_{n=1} \frac{(-1)^n}{n!}  \frac{d^{n-1}}{d t^{n-1}} 
\big( H \pi^n\big)  \, .
\ea
Our job of obtaining the interaction Hamiltonian at any order is now complete. Eq. (\ref{HI-eq2}) provides the non-perturbative form of the Hamiltonian in terms of  $\pi$  while $\calR$ itself is non-linearly related to $\pi$ via 
Eq. (\ref{R-eq}). \\


{\bf Higher Order Loop Corrections:} 
The general form of the Hamiltonian 
(\ref{HI-eq2}) with the non-linear relation (\ref{R-eq}) enables us to calculate the loop corrections in the power spectrum at any order. Here we calculate the $L$-loop corrections in the setup considered in \cite{Kristiano:2022maq}. This is a three-stage model $\mathrm{SR\rightarrow USR\rightarrow SR}$ which is employed for PBHs formation. The USR phase is limited to the period $\tau_s \leq \tau \leq \tau_e$ during which the power spectrum experiences rapid growth. We have an instant transition from the USR phase to the final SR phase but the evolution of the mode function can be either sharp or mild which
is controlled by the relaxation parameter $h$. The case of one-loop correction was studied in \cite{Firouzjahi:2023aum} with the effects of both cubic and quartic Hamiltonians included.   

\begin{figure}[t]
	\centering
	\includegraphics[ width=1\linewidth]{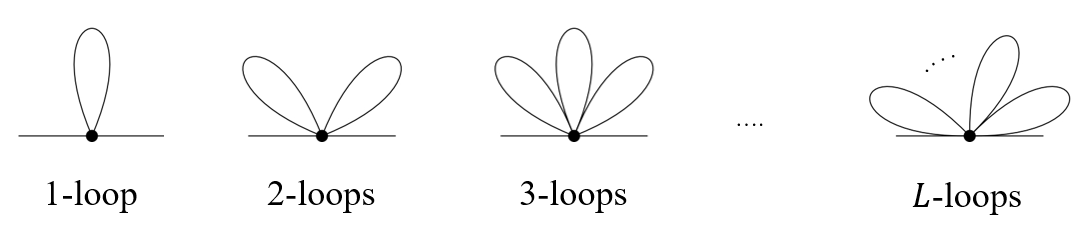}
	\caption{ The  one-particle irreducible Feynman diagrams with  one vertex 
	for various values of $L$.    }
\label{diagram}
\end{figure}

As $L$ increases the number of one-particle irreducible Feynman diagrams increases rapidly. Many of these diagrams involve multiple vertices which are 
difficult to handle for the in-in analysis. The reason is that multiple vertices yield multiple nested time integrals with non-trivial mode functions which in general are  complicated to calculate analytically. However, among all of these Feynman diagrams, there is one particular diagram containing a single vertex with $L$ loops attached to it as depicted in Fig. \ref{diagram}.  The in-in integral for this diagram  involves a single time integral which can be taken analytically. Besides this technical simplicity, this diagram has the dominant contribution in loop correction as we shall elaborate later on. Having said this, we comment that the analysis even for this single vertex case is non-trivial.


To calculate the $L$-loop correction, we need the interaction Hamiltonian at the 
order $n=2L+2$, which from Eq.~\eqref{HI-eq2}, takes the following form,
\ba
\label{H-int}
{\bf H}_{2L+2}=  \frac{M_P^2 H^2}{(2 L)!} \int d^3 x   a^3 \Big[ {\tilde\epsilon}^{(2 L)}   
\dot \pi^2 + \epsilon^{(2 L)}  \partial_i\pi \partial^i \pi \Big] \pi^{2L}. 
\ea
In our notation, $X^{(k)}$ stands for the $k$-the order derivative, i.e. 
$X^{(k)}(z) \equiv \frac{d^{k}}{d z^{k}} X(z)$ and ${\tilde\epsilon}^{(2 L)}$ is defined via,  
\ba
{\tilde\epsilon}^{(2 L)} \equiv \epsilon(t)^2 \Big(\frac{1}{\epsilon(t)}\Big)^{(2 L)}\, .
\ea

The measurement of the power spectrum is made at the time of the end of inflation 
$\tau_0$ in the final attractor phase in which the derivatives like $\dot \pi, \ddot \pi$ and $\dot H, \ddot H$ etc are all suppressed. In this limit, the non-linear relation (\ref{R-eq})  between $\calR$ and $\pi$ is simplified to a linear one, $\calR = -H \pi + {\cal O} (\epsilon^2)$. Correspondingly,  the $L$-loop correction associated with the one-vertex diagram at the end of inflation $\tau_0$ is given by 
\ba
 \label{in-in-int}
 \langle \calR_{\bfp_{1}} \calR_{\bfp_{2}} \rangle
= - 2 H^2 \mathrm{Im}  \int_{-\infty}^{\tau_0} d \tau   \big \langle {\bf H}_{2L+2} (\tau) \pi_{\bf p_1} \pi_{\bf p_2}(\tau_0)  \big\rangle.
\ea

There are three different types of contributions in the above integral. The first contribution is from the bulk of USR in which $\tau_s< \tau <\tau_e $. The second type is from the boundary at $\tau=\tau_e$ due to jump in $\eta$ which is encoded in quantities such as $\dot \eta, \ddot \eta $ etc, appearing in  $\epsilon^{(2L)}$ and $\tilde \epsilon^{(2L)}$. The third type of contribution comes from the final SR phase, $\tau > \tau_e$. However, one expects that these contributions to be subleading in our setup with a short relaxation period  with $|h| > 1$ in which the mode function rapidly approaches to its attractor value. We have verified this conclusion, see also 
\cite{Firouzjahi:2023aum} for  one-loop case.  

Below we calculate loop corrections from the bulk  and  
from the boundary at $\tau_e$. 
Note that since the mode functions are not amplified during the first SR phase, the loop corrections from the boundary terms at  $\tau_i$ are negligible compared to those from the bulk and from  $\tau_e$. 
The fractional loop corrections from the bulk 
is obtained to be, 
\begin{equation}
\label{Bulk-cont}
\frac{\Delta \calP_{\mathrm{Bulk}} }{\calP_{\mathrm{cmb}}}\Big|_{L-\mathrm{loop}}=
R_{\mathrm{B}}(L)  \big( \Delta N \calP_e \big)^L \, ,
\end{equation}
where $\calP_{\mathrm{cmb}}$ is the CMB power spectrum and $\calP_e$
 is the power spectrum at $\tau=\tau_e$, related 
to $\calP_{\mathrm{cmb}}$ via 
\ba
\label{Pe}
\calP_e = \frac{H^2}{8 \pi^2 M_P^2 \epsilon_e} =  \calP_{\mathrm{cmb}} e^{6 \Delta N}\, .
\ea
In addition, the numerical prefactor  $R_{\mathrm{B}}(L)$ is given by, 
\ba
R_{\mathrm{B}}(L)  \equiv \Big[ {12 L+ h} + \frac{1}{\Delta N}\frac{L  (12 L-h-4)}{2 (6 L-5) (3 L-1) } \Big] 
\frac{18^L}{h \Gamma(L+1)}. \nonumber
\ea
For large $L$,  $R_{\mathrm{B}}(L) $ behaves as,
\ba
\label{RB-assym.}
R_{\mathrm{B}}(L) \rightarrow \frac{6 \sqrt{2 L} }{h \sqrt{\pi}} \big( \frac{18 e}{L}\big)^L  \, , \quad \quad (L\gg 1) \, .
\ea
This indicates that the loop corrections from the bulk decreases for 
$L\gg1$ and the total loop corrections from the resummation of all $L$-loop corrections should be converging. Interestingly, the 
resummation of $L$-loop corrections can be made, yielding \footnote{We use Maple and Mathematica softwares to perform the analytical and numerical analysis.}
\begin{widetext}
\ba
\label{resum}
\frac{\Delta \calP_{\mathrm{Bulk}} }{\calP_{\mathrm{cmb}}}
\Big|_{\mathrm{resummed}}
=-1 + {e^{18 \Delta N \calP_e} } 
\big( 1+ \frac{216}{h} \Delta N \calP_e \big) 
 -\Big(\frac{9 (h-8)}{2 h } \calP_e\Big) \,  {_3}\mathrm{F}_{3} \Big(\Big[\frac{1}{6}, \frac{2}{3}, \frac{20-h}{12} \Big],  \Big[\frac{7}{6}, \frac{5}{3}, \frac{8-h}{12} \Big], 18 \Delta N \calP_e \Big)
\ea
\end{widetext}
in which $ {_3}\mathrm{F}_{3}$ is the generalized hypergeometric function. 
The resummed loop correction  Eq. (\ref{resum}) is interesting on its own right. This may be compared to the resummation of the IR diagrams in finite temperature field theory. This similarity may be invoked from the fact that the dS spacetime is a thermal background with the Hawking temperature $\frac{H}{2 \pi}$. While the  hypergeometric function is not insightful, but  the dominant contribution comes mostly from the first two terms,  scaling like ${e^{18 \Delta N \calP_e} }$. With  $\calP_e$ given in Eq. (\ref{Pe}), we conclude that  the resummed loop correction will quickly go out of perturbative control for large enough $\Delta N$.

The in-in analysis for the boundary terms at $\tau=\tau_e$ proved to be non-trivial. This is because  the terms $\epsilon^{(2L)}$ and $\tilde \epsilon^{(2L)}$ involve various powers of $\delta (\tau-\tau_e)$ and its derivatives.  
Performing the integrals, one finds 
\cite{longpaper}, 
\ba
\label{boundary-cont}
\frac{\Delta \calP_{\mathrm{boundary}} }{\calP_{\mathrm{cmb}}}\Big|_{L-\mathrm{loop}}=
R_{\mathrm{b}}(L) \big( \Delta N  \calP_e\big)^L \, ,
\ea
in which  $R_{\mathrm{b}}(L)$ has a complicated form.  Unlike 
$R_{\mathrm{B}}(L)$ from the bulk, $R_{\mathrm{b}}(L)$ grows rapidly for large
$L$, 
\ba
\label{Rb}
R_{\mathrm{b}}(L)  \rightarrow  ( 18 L)^L   \quad \quad (L\gg 1) \, .
\ea
As a result, for large values of $L$ the loop correction from the boundary  scales like $\big(L \Delta N e^{6 \Delta N}  \calP_{\mathrm{cmb}}\big)^L$. 
This indicates that the loop corrections from the boundary are far more important than those of bulk and will go out of control for large $L$. This is because the boundary terms are generated by localized sources such as $\dot \eta, \ddot \eta, \dddot \eta, \dot \eta^2$ etc which involve $\delta (\tau-\tau_e)$ and its derivative to various orders. The larger is $L$, the higher is the order of derivatives $\ddot\eta, \dddot \eta$ etc  or the powers  $\dot \eta^2, \dot \eta^3$ etc  which in turn  induce larger corrections. 

Adding the contributions of the bulk and boundary, the  total fractional loop correction at the $L$-loop level is given by,
\ba
\label{total-loop}
\frac{\Delta \calP_{\mathrm{tot}} }{\calP_{\mathrm{cmb}}}\Big|_{L-\mathrm{loop}} &=&
\big( R_{\mathrm{b}}(L) + R_{\mathrm{B}}(L) \big)  
\big( \Delta N \calP_e\big)^L \\
\label{asym}
 &\sim& \big( 18 L  \Delta N e^{6 \Delta N} \calP_{\mathrm{cmb}}\big)^L  
 \quad  (L \gg 1). 
\ea

A few important conclusions can be drawn from the above result. First, we see the strong sensitivity of the loop corrections to $\Delta N$, the duration of the USR phase.  Furthermore, for any given value of $\Delta N$ there is a threshold value $L_c$  in which  the fractional loop correction becomes unity and rapidly increases beyond the  perturbative control for $L> L_c$. From Eq. (\ref{asym}), $L_c$ is estimated as 
\ba
 L_c \simeq \Big( 18  \Delta N e^{6 \Delta N} \calP_{\mathrm{cmb}}\Big)^{-1}. 
\ea
This indicates that $L_c$ is inversely and exponentially related to $\Delta N$ so  the larger is $\Delta N$, the smaller is $L_c$. For example, in standard scenarios of PBHs formation \cite{Garcia-Bellido:2017mdw, Germani:2017bcs, Biagetti:2018pjj}, one requires $\Delta N \simeq 2.5$ in order to increase the power spectrum to the required level $\calP_e \sim 10^{-2}$. This yields  $L_c \simeq 3.4$, meaning that the perturbative loop corrections will be lost  at the four-loop level. However, we caution that this value for 
$L_c$ is obtained in our simplified picture where the transition from the 
USR to SR is (a): instantaneous and (b): sharp. In the realistic situation that the transition may not satisfy either of the conditions (a) or (b) above, then the analysis is more complicated and the specific conclusion that the perturbativity is lost at $L=4$ has to be relaxed.

Finally, from the asymptotic  $L^L$ dependence of the loop corrections in Eq. (\ref{asym}), one concludes that the one-vertex diagrams as shown in Fig. \ref{diagram} have the leading loop corrections among all other diagrams, see also  \cite{Leblond:2010yq}. 
As an example of multi-vertices Feynman diagrams, consider a one-particle irreducible   diagram with two vertices, one vertex of quartic Hamiltonian ${\bf H}_4$ and another vertex of ${\bf H}_{2 L}$ in which the external lines are both attached to 
${\bf H}_4$ vertex.  This  corresponds to replacing  $R_b(L)$ by $R_b(L-1)$ in the in-in integrals. From the asymptotic form of $R_b(L)$ in Eq. (\ref{Rb}), one concludes that  the contribution of 
this two-vertices diagram is smaller than that of our one-vertex diagram  by a factor $\big(18L\big)^{-1}$. On the other hand,  there are many multi-vertices diagrams  having loop corrections of either signs. So the sum of their  contributions may be smaller than that of our one-vertex diagrams as well.

\begin{figure}[t!]
	\centering
	\includegraphics[ width=1\linewidth]{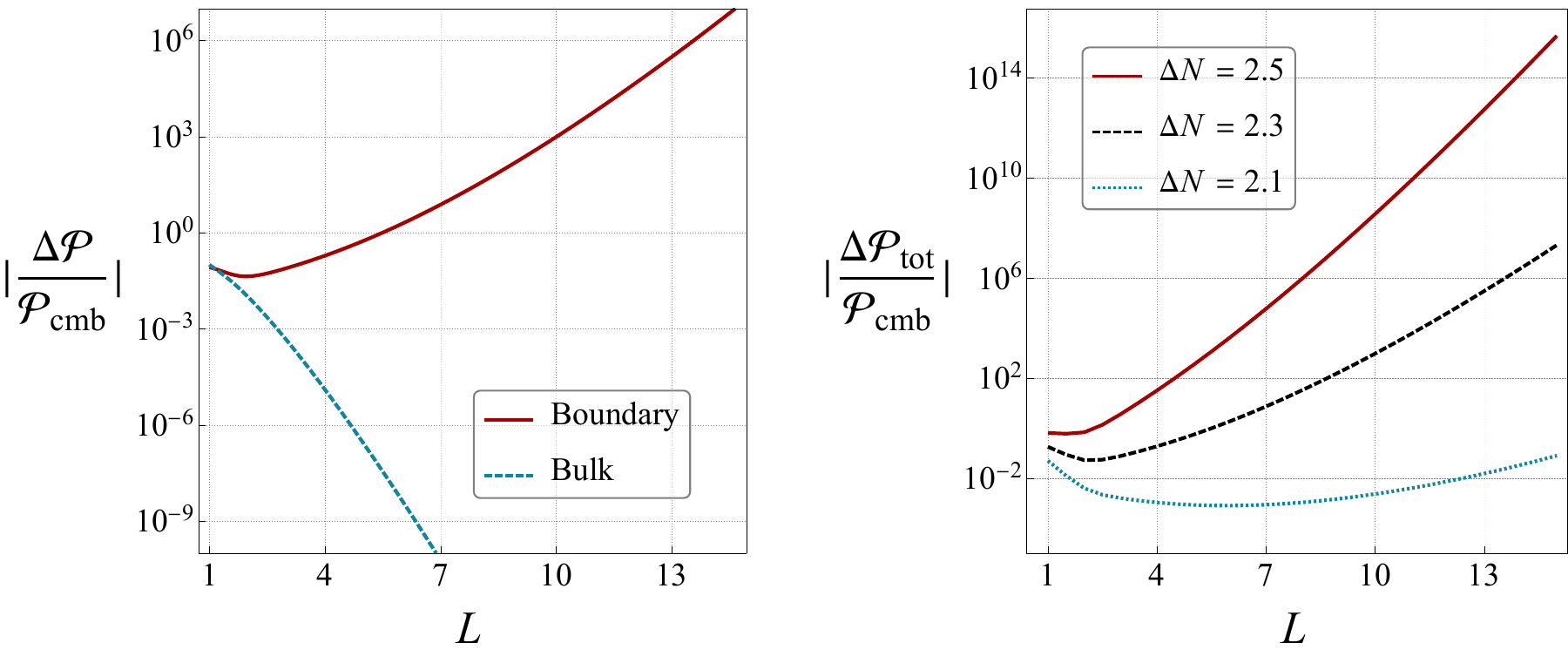}
	\caption{ Loop corrections with $h=-6$.  
	{\bf Left:} The contributions of the bulk Eq. (\ref{Bulk-cont}) (dashed curve) and the  boundary Eq. (\ref{boundary-cont}) (solid curve) with $\Delta N=2.3$.  As $L$ increases, the bulk contributions fall off while those of boundary increase. 
{\bf Right:} Total loop corrections Eq. (\ref{total-loop}) for different  values of $\Delta N$: $\Delta N=2.1, \Delta N=2.3$ and $\Delta N=2.5$, from bottom to top respectively. We see the strong dependence on $\Delta N$ while $L_c$ is smaller for larger values of $\Delta N$.    }
\label{fig2}
\end{figure}


In Fig. \ref{fig2} we have presented the loop corrections in terms of $L$.   
In the left panel we have compared the contributions of 
 bulk with those of the boundary. It clearly shows that the loop corrections from the boundary dominate over the bulk. In the right panel, the total loop corrections from  different values of $\Delta N$ are shown. This plot shows the exponential dependence of loop corrections 
 on $\Delta N$ and also confirms the conclusion that 
larger values of $\Delta N$ yield to smaller values of $L_c$. \\

{\bf Conclusions:} In this work, employing the formalism of EFT of inflation, 
we have calculated the action and the Hamiltonian in USR setup to all order in perturbation theory in the decoupling limit. Interestingly, we were able to find a compact non-perturbative expression for the Hamiltonian in terms of the Goldstone field $\pi$. We also obtained a non-linear relation between $\calR$ and  $\pi$. 
 As a non-trivial application, we have calculated the higher order loop corrections in the setup usually employed for PBHs formation,
{i.e. the transient USR model with a short relaxation time to the attractor phase}.  Our results show that the loop corrections become significant for large $L$. The dominant contributions come from the boundary terms localized at $\tau=\tau_e$. This is originated from the singular behaviours of the contributions like 
$\dot \eta, \ddot \eta, \dot \eta^2$ etc which involve $\delta (\tau-\tau_e)$ and its derivatives to higher order. 

Motivated by these results, we conjecture that loop corrections may become important in other models of inflation in which the potential involves localized features which are sharp enough to be  modelled by a delta function. 
The resultant loop corrections are model-dependent but one may expect the $L$-loop
correction to be in the  form  $R_f(L) \calP_f^L $  in which $\calP_f$ is the power spectrum  at the onset of feature and the parameter $R_f(L)$ is similar to our functions $R_b(L)$.  Finally, in this work we did not study the renormalization of the loop corrections. This was studied for the case of one-loop in 
\cite{Sheikhahmadi:2024peu}. While renormalization is an important step to read off the physical results, but one expects that the above conclusions about the magnitude of loop corrections to remain valid even when  the effects of renormalization are taken into account.  This is because there is no underlying symmetry to cancel the loop corrections order by order, as shown explicitly in the case of one-loop in  \cite{Sheikhahmadi:2024peu}.\\

{\bf Acknowledgment:} We are grateful to A. A.  Abolhasani, 
M. H. Namjoo, A. Riotto and    H. Sheikhahmadi for valuable discussions and comments. 
The work of H. F. is  partially supported  by INSF of Iran under the grant number 4038049. B. N. thanks ICTP for hospitality where this work was in progress.  
\appendix

\bibliography{short-paper}{}

\end{document}